\pdfoutput=1
\documentclass{elsarticle}
\usepackage{amsmath,breqn}

\usepackage{graphicx}
\usepackage{tabularx}
\usepackage{float}

\begin{document}

\begin{frontmatter}

\title{Using Stein's Unbiased Risk Estimate (SURE) to Optimize Level of Decomposition in Stationary Wavelet Transform Denoising}

\author{Md Yusof, M.R. and Ariffin, A.K. }

\address{Dept. of Physics, Faculty of Science and Mathematics, Universiti Pendididkan Sultan Idris}


\begin{abstract}
A method of determining the optimum number of levels of decomposition in soft-thresholding wavelet denoising using Stationary Wavelet Transform (SWT) is presented here. The method calculates the risk at each level of decomposition using Stein’s Unbiased Risk Estimate (SURE), analogous to calculating the sum square error of the denoising process. The SURE risk is found to reach minimum at the same level of decomposition as the sum square error. The advantage of this method is that the clean signal need not be known a priori. The method can be used with either a priori known noise variance or an estimate of the noise variance. This allows the determination of the optimum level of decomposition for wavelet denoising of an unknown signal so long as the noise variance can be estimated.
\end{abstract}
\end{frontmatter}


%


\section{Introduction}
\subsection{Literature Review}
Wavelet denoising has been proposed for removal of noise from a signal in many works \cite{ShafriMather2005}\cite{ShafriYusof2009}\cite{YangJudd2004}. Schmidt and Skidmore \cite{SchmidtSkidmore2004} used a very basic method of wavelet denoising for vegetation reflectance spectrum and determined the optimum level of decomposition by visually inspecting the resulting signal. This was the first such use of stationary wavelet transform denoising in the field of vegetation reflectance spectroscopy; previous works used convolution with low-pass kernels such as Savitsky-Golay filters which was adequate when dealing with relatively low spectral resolutions \cite{RuffinKing1999}. 

With the advent of newer data acquisition systems capable of acquiring high resolution spectra, the level of noise can be expected to increase as an implication of the Nyquist Sampling Theorem, where the highest frequency component present in a signal is half of the sampling frequency. Higher sampling leads to more high frequency noise. More efficient denoising strategies are needed. This is especially if differential analysis is used to accentuate features of the signal \cite{ShafriMather2005}. As the differential operation acts as a high-pass filter, it will accentuate the high frequency noise.

Others have used wavelet denoising for processing astronomical signals \cite{StarckMurtagh2001}\cite{Kolaczyk96}\cite{Sulochana2016}, medical imaging \cite{RafieeRafiee2011,NowakBaraniuk1999}, spectroscopy data and hyperspectral imaging \cite{Pizurica2002, ShafriMather2005, ShafriYusof2009}. These present the additional challenge of dealing with signals with non-constant noise variance, as the noise follows Poisson statistics. For large photon counts, the variance is proportional to the photon count. The simplest approach is to apply a variance-stabilizing transform such as the Anscombe Transform, prior to applying thresholding rules that were originally designed for signals with Gaussian noise with constant noise variance \cite{MakitaloFoi2013}.
In the case of astronomical signals and similar low photon intensity signals, a further constraint is the typically low signal amplitude due to the vast distance of the source. The Anscombe Transform approach may not suffice \cite{MakitaloFoi2013}.

The basic principle of most wavelet denoising techniques is to modify the wavelet coefficients obtained by the transform prior to reverse transforming it. This is usually accomplished by some thresholding rule. A method called “soft thresholding” where wavelet coefficients are thresholded and the remaining coefficients reduced by the value of the threshold prior to reconstruction has been proposed \cite{Donoho1995}. This was performed using a relatively straightforward threshold, the Universal Threshold $\sigma\sqrt{2 \log⁡L }$ where L is the length of the data. Donoho and Johnstone developed the idea further with newer thresholding rules: MiniMaxi, SUREShrink and Hybrid SURE \cite{Mallat1999}. 

Wavelet transform algorithms that are commonly mentioned are the Continuous Wavelet Transform (CWT), the Mallat Algorithm, and the Stationary Wavelet Transform (SWT)\cite{Pizurica2002}. The Stationary Wavelet Transform (SWT) is one of the algorithms used for implementing the discrete wavelet transform, for denoising. Of several translation-invariant schemes tested, the SWT was found to yield the best result in terms of RMSE \cite{CoifmanDonoho1995}. Newer works have mentioned the Lifting Algorithm, also known as Second Generation Wavelet \cite{EbadiShafri2014} and its use in denoising. The advantages include not needing the length of the signal to be a power of two. 

Each algorithm has distinct advantages and disadvantages. The Mallat Algorithm is fast to compute, for the amount of data to be processed is halved at each level of decomposition for 1-D data, quartered for 2-D data \cite{Mallat1999}. The SWT is translation-invariant, the signal features do not shift \cite{Pizurica2002}. Furthermore the SWT enables the use of wavelets with short support lengths which can retain small signal features better. This was demonstrated by Schmidt and Skidmore and their use of the Haar wavelet, which is only two points in length \cite{SchmidtSkidmore2004}. The Lifting Algorithm is said to be excellent in denoising extremely noisy data \cite{Ebadi2013}.

\subsection{Optimization of Parameters in SWT Wavelet Denoising}
Wavelet denoising requires fine-tuning of several parameters: the number of levels of wavelet decomposition, the wavelet function used and the thresholding rule. One method involves creating a simulated clean signal, and adding Gaussian White Noise \cite{SinghTiwari2006}. The MSE was calculated between the simulated signal and the product of denoising to determine the optimum parameters. Singh and Tiwari used it to determine the optimal thresholding rule for ECG signals and to verify that the correlation method selected the optimum wavelet. This method was further developed in \cite{RafieeRafiee2011} for medical imaging.

We propose an alternative approach to determine one of the parameters to be optimised: the optimum level of decomposition. We will fix the thresholding rule and wavelet by using the “Universal Threshold” \cite{Donoho1995} and the Daubechies db4 wavelet, in order to focus on determining the optimum level of decomposition. 

In this work we will use the SWT algorithm. The principles are well understood and can be expressed here as a series of convolutions of the high-pass and low-pass decomposition and reconstruction kernels. This lends it to being expressed as a convolution, and so as a linear matrix operation, which lends it to expression as a term in Stein’s Unbiased Risk Estimate.

The determination of the optimum number of levels of decomposition is scarcely studied. The most well understood approach used MSE with a simulated clean signal with additive noise \cite{SinghTiwari2006}. Others have relied on visual inspection, noting that performing wavelet denoising at excessive levels of decomposition caused the signal to deteriorate \cite{SchmidtSkidmore2004}, or stopped at two levels of decomposition as a rule of thumb \cite{Sulochana2016}. 

The mean square error (MSE) method requires that a clean signal with characteristics similar to the one we are studying be simulated. Noise is introduced to this signal, and a denoising process is performed on this simulated signal. The parameters of the denoising process are adjusted until the MSE is minimized. The procedure is then applied to the signal that we are actually studying with the assumption that what is optimum denoising for the simulated signal will also be optimal for the actual unknown signal. For wavelet denoising, the parameters to be adjusted are the wavelet, threshold and number of levels of decomposition to be used in denoising. 

An alternative approach is presented in \cite{Stein1981}, using Stein’s Unbiased Risk Estimate (SURE). Stein’s Lemma enables the risk to be calculated without the clean signal being known. It does, however, require that the variance of the noise be known. With this the risk can be calculated in order to determine the optimum value of the parameters to be used in wavelet denoising. It also requires the denoising process to be expressed as an estimator in order to be included as a term in SURE.

In this work we will make use of SURE to determine the optimum level of wavelet denoising, when the Universal Threshold Rule is applied with the Stationary Wavelet Transform. We will express the SWT wavelet denoising algorithm as a term in Stein’s lemma and calculate the risk, in order to determine the optimum number of levels of decomposition. The result will be compared to the MSE.
SURE is a method that can be used as a surrogate to the mean square error method, which makes it useful for when the clean signal is not known. 

\section{Theory}
\subsection{Wavelet Transform}

The Continuous Wavelet Transform (CWT) is given by 

\begin{equation}
[W_\psi f](a,b)=\frac{1}{\sqrt(a)} \int_{-\infty}^\infty \psi\left(\frac{x-a}{b}\right)f(x) dx
\end{equation}

The Wavelet Transform is a convolution between a signal f(x) and a finite-length wavelet function $\psi$ of scale a. The result is a set of wavelet coefficients W, the values of which depend on the scale a of the wavelet.

The CWT is said to be redundant since the wavelet coefficients generated encode high and low frequency components. Other wavelet algorithms employing pairs of high and low pass filters, have been devised, with the high-pass filters encoding the high frequencies to produce wavelet detail coefficients while the low-pass filters exclude precisely those frequencies to produce the wavelet approximation coefficients. The filters are dilated and the process is repeated on the approximation coefficients. Reconstructing the signal from the coefficients employ a pair of high and low-pass reconstruction filters which are time-reflected versions of the decomposition filters. The four filters together make up the quadrature mirror filters.

Two very well-known algorithms that implement wavelet transform using the quadrature mirror filters are the Mallat Algorithm and the Stationary Wavelet Transform (SWT). The SWT has some advantages over the Mallat Algorithm such as permitting the use of short support-length wavelets for denoising \cite{SchmidtSkidmore3504}, and being translation-invariant. While newer algorithms such as second-generation wavelets and the Lifting Algorithm have become the cutting edge \cite{Ebadi2013}, the SWT still has the advantage of being well established and understood.

The Stationary Wavelet Transform (SWT) uses of a pair of convolution kernels, the high-pass wavelet function $\psi$ and the low-pass approximation function $\phi$. Convolving the signal with $\psi$ and $\phi$ results in the sets of detail coefficients and approximation coefficients $d_1$ and $a_1$ respectively. The two convolution kernels are then dilated; which in SWT is done by dyadic upsampling. The dilated kernels are applied to the approximation coefficients generated by the preceding level to generate detail and approximation coefficients $d_2$ and $a_2$. The process is repeated for $N$ levels \cite{Mallat1999}.

This process is sketched out for three levels in Figure 1, where $\psi_i$  and $\phi_i$ represent the normalized high-pass and low-pass kernels respectively and $\psi_{i+1}$ and $\phi_{i+1}$ are the dyadically upsampled kernels of $\psi_i$ and $\phi_i$.

\begin{figure}[H]
\includegraphics[width=600pt, trim={50pt 300pt 200pt 0},clip]{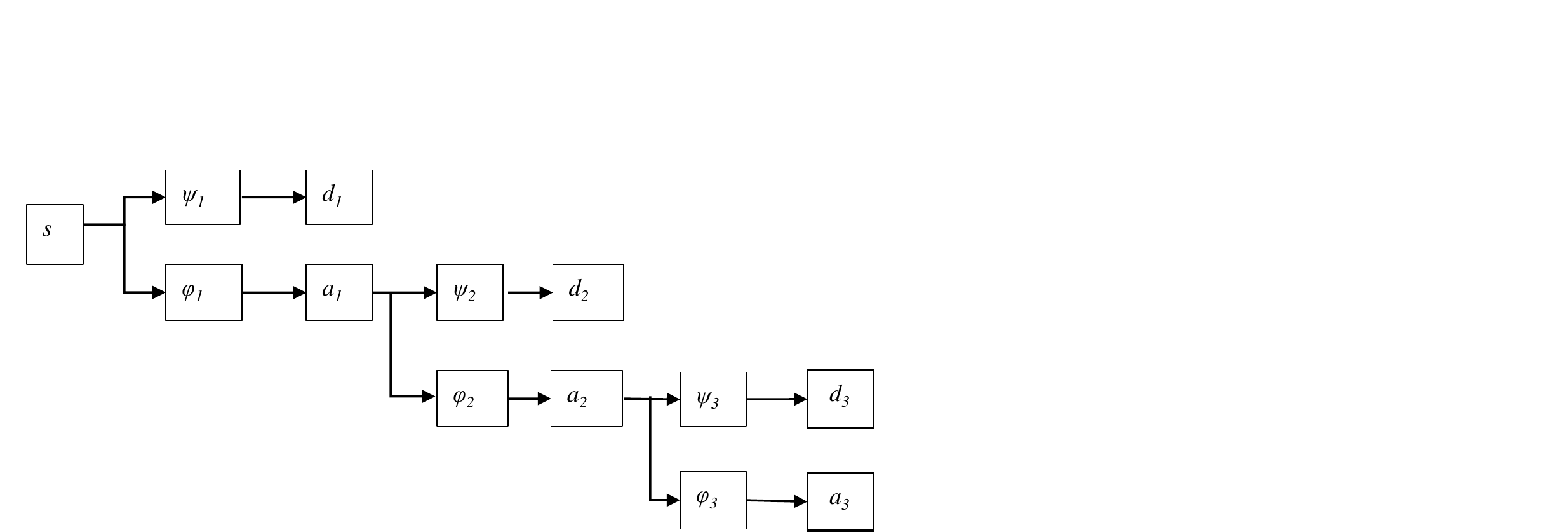}
\caption{SWT for three levels of decomposition}
\end{figure}

For three levels of wavelet decomposition, the detail coefficients $d_1$, $d_2$ and $d_3$ can be shown to be
$$d_1=\psi_1*s$$
$$d_2=\psi_2*\phi_1*s$$
$$d_3=\psi_3*\phi_2*\phi_1*s$$
while the approximation coefficients
$$a_3=\phi_3*\phi_2*\phi_1*s$$
where $\psi_{i+1}=\psi_i [\uparrow2]$ and $\phi_{i+1}=\phi_i [\uparrow2].$

Reconstructing the original signal is done by performing the inverse wavelet transform by convolving the detail and approximation coefficients using the wavelet reconstruction kernels $\phi_1',\phi_2',\hat{phi_3}'$ and $\psi_1',\psi_2',\psi_3'$, which are the time-reversed versions of the decomposition kernels, where likewise $\psi_{i+1}'=\psi_i' [\uparrow2]$ and $\psi_{i+1}'=\psi_i'[\uparrow2]$.

$$s=\psi_1'*d_1+\phi_1'*\psi_2'*d_2+\phi_1'*\phi_2'*\phi_3'*a_3$$
For N levels, the reconstruction of the signal from the detail and approximation coefficients can be written

\begin{equation}\label{eq: wavelet reconstruction}
s=\psi_1'*d_1+\sum_{j=2}^N \left(\prod_{i=1}^j \phi_i'*\right)\psi_j'*d_j+\prod_{j=1}^N(\phi_j'*)a_N
\end{equation}

where
\begin{equation} \label{eq: detail coeff 1}
d_1=\psi_1*s
\end{equation}

\begin{equation}\label{eq: detail coeff j}
d_j=\left(\prod_{i=1}^{j-1}\phi_i\right)*\psi_j*s
\end{equation}

\begin{equation}\label{eq: approx coeff}
a_N=\prod_{i=1}^N(\phi_i'*\phi_i)*s 
\end{equation}

Here we “borrowed” the symbol $\prod$, used for serial multiplication, to signify serial convolution.

\subsection{Wavelet Denoising by Soft Thresholding}

Wavelet denoising involves modifying the wavelet detail coefficients by certain rules, such that the signal, when reconstructed, has been denoised. Thresholding of the wavelet detail coefficients is a commonly used method. The VisuShrink wavelet thresholding rule is used here due to its simplicity (Donoho and Johnstone 1994).
\begin{equation}
\tau=\sigma\sqrt{2 \log L}
\end{equation}
where L is the length of the data sample and σ is the standard deviation of the noise. The detail coefficients d1 consist mainly of noise and the standard deviation of the d¬1 coefficients are taken to be the that of the noise. The MAD (median absolute deviation) may be used as an intermediate step for obtaining the standard deviation [17].

$$\sigma=MAD*1.4826$$

The wavelet detail coefficients are thresholded such that:
\begin{equation}
d_i=0 \text{	if	} |d_i |<\tau
\end{equation}

For soft thresholding [14] the wavelet detail coefficients are further modified by reducing them by the value of the threshold:
\[
  \hat{d_i}=\begin{cases}
               d_i=0 & \text{if	}  |d_i|<\tau\\
               d_i=d_i- sign(d_i)\tau &\text{if } |d_i| \geq \tau\\

            \end{cases}
\]

After the coefficients have been thresholded, the signal is reconstructed by the inverse wavelet transform. The thresholding rule is supposed to eliminate coefficients that contain only noise. The VisuShrink threshold rule is supposed to be the upper limit of noise in a signal. 

\paragraph{Proposition 1}
The denoising process of a signal s using soft thresholding can be written 

\begin{multline}
\label{eq:wavelet_convolution}\hat{s}=\delta_1(\psi_1' *\psi_1*s)+\sum_{j=2}^N\left(\delta_j\prod_{i=1}^{j-1}(\phi_i'*\phi_i)*(\psi_j'*\psi_j)\right)*s+\prod_{j=1}^N(\phi'_j*\phi_j*)s  \\  -\psi'_1*\delta_1\epsilon_1\tau   -\sum_{j=1}^N\left(\prod_{i=1}^{j-1}\phi_i'*\right)\psi_j'*\delta_j\epsilon_j\tau
\end{multline}

where

\[
\delta_i(x)=\begin{cases}
	0&\text{if }|d_i(x)|<\tau\\
	1&\text{if }|d_i(x)|\geq\tau
	\end{cases}
\]

and

\[
\epsilon_i(x)=\begin{cases}
	-1&\text{if }|d_i(x)|<0\\
	1&\text{if }|d_i(x)|>0
	\end{cases}
\]	

Proof:
The soft thresholding operation is written:
$$s=\psi_1'*\hat{d_1}+\sum_{j=2}^N \left(\prod_{i=1}^j \phi_i'*\right)\psi_j'*\hat{d_j}+\prod_{j=1}^N(\phi_j'*)a_N$$

where
$$\hat{d_j}=\delta_j\times(d_j-\epsilon_j\tau)$$

Thus

$$s=\psi_1'*\delta_1\times(d_1-\epsilon_1\tau)+\sum_{j=2}^N \left(\prod_{i=1}^j \phi_i'*\right)\psi_j'*\delta_j\times(d_j-\epsilon_j\tau)+\prod_{j=1}^N(\phi_j'*)a_N$$

Combining with \ref{eq: detail coeff 1}, \ref{eq: detail coeff j}, \ref{eq: approx coeff}, we obtain \ref{eq:wavelet_convolution}:

$$\hat{s}=\delta_1(\psi_1' *\psi_1*s)+\sum_{j=2}^N\delta_j\prod_{i=1}^{j-1}(\phi_i'*\phi_i)*(\psi_j'*\psi_j)*s+\prod_{j=1}^N(\phi'_j*\phi_j*)s$$    
$$-\psi'_1*\delta_1\epsilon_1\tau    -\sum_{j=1}^N\left(\prod_{i=1}^{j-1}\phi_i'*\right)\psi_j'*\delta_j\epsilon_j\tau$$

\paragraph{Proposition 2} We can define $A$ and $B$ so that the wavelet denoising becomes in effect a convolution operation:

\begin{equation}
\hat{s}=A*s +B
\end{equation}

where

\begin{equation}\label{eq: A}
A=\delta_1(\psi'_1*\psi_1)+\sum_{j=2}^N\delta_j\prod_{i=1}^{j-1}(\phi'_i*\phi_i*)(\psi'_j*\psi_j)+\prod_{j=1}^N(\phi'_j*\phi_j*)
\end{equation}

and 

\begin{equation}\label{eq: B}
B=-\psi'_1*\delta_1\epsilon_1\tau    -\sum_{j=1}^N\left(\prod_{i=1}^{j-1}\phi_i'*\right)\psi_j'*\delta_j\epsilon_j\tau
\end{equation}

The term $\prod_{j=1}^N(\phi'_j*\phi_j*)$  in \ref{eq: A} creates a low-pass filter kernel that increases in width with the number of levels of decomposition used. The low pass kernel filters noise from the signal but tends to oversmooth regions of the signal with sharp edges. 

At low levels of wavelet decomposition, the first two terms of \ref{eq: A} consist of high frequency components which are thresholded. For smooth regions of the signal, the detail coefficients are completely thresholded and the first two terms of \ref{eq: A} do not contribute to the equation. At sharp edges of the signal, the detail coefficients are nonzero and the first two terms contribute and add high frequency components to the smoothing kernel. The wavelet denoising process can be thought of as a convolution between a signal and a low pass filter that is interrupted at sharp signal edges. 

As higher levels of decomposition are used, the detail coefficients begin to encode larger features of the signal with low frequency components instead of just sharp signal features. Thresholding these coefficients leads to signal features being lost and a distorted reconstructed signal. It is the purpose of this work to determine the optimum number of levels of decomposition when using the VisuShrink thresholding rule for denoising, in order to use SWT wavelet denoising without creating excessive distortion.

The B terms in \ref{eq: B} are a correction against pseudo-Gibbs phenomenon, which occur at sharp edges of the reconstructed signal. The first term at the first level of decomposition, $-\psi'_1*\delta_1\epsilon_1\tau$ translates $\psi'_1$ to position $x$, where $\epsilon_1$ determines the sign and τ is a scalar. At sharp features this term, as well as the second term of B, will reduce the pseudo-Gibbs phenomenon.

\subsection{Stein's Unbiased Risk Estimate (SURE)}

Evaluation of a noise-removal process is often done by performing a MSE, RMSE or SSE (sum square error) analysis. A known clean signal, usually generated from a model, is treated with noise with known variance and the noise-removal process with certain parameters is performed. The result is evaluated by the mean square error between the known clean signal and the signal post-operation. Having a procedure that performs satisfactorily, minimizing the MSE, the procedure is then used with those parameters on a real world signal whose clean form is unknown, but has characteristics similar to the model signal. This was the method used by researchers such as Singh and Tiwari, etc. \cite{SinghTiwari2006}.

Stein's Unbiased Risk Estimate (SURE) provides a surrogate to the MSE, where the clean signal is unknown, but the variance of the noise can be determined or is known a priori. SURE calculates a risk value that stands in for the MSE. The parameters for treating a noisy signal can be adjusted to minimize the SURE risk, which can stand for minimizing the MSE risk as well.
SURE is written 

\begin{equation}
SURE=-n\sigma^2+||\hat{s}-s||^2+2\sigma^2\sum_{i=1}^n\frac{\partial\hat{s}}{\partial{s_i}}
\end{equation}

where $\sigma$ is the standard deviation of the noise, $s$ is the unprocessed unknown clean signal, and $\hat{s}$ the signal after the noise-removing procedure \cite{Stein1981}\cite{GubbiSeelamantula2014}.

The expectation value of the SURE risk is equal to the expectation value of the sum squared error:
\begin{equation}
E(SURE)=E(||\hat{s}-s_0 ||^2 )=MSE
\end{equation}

where $s_0$ is the unknown clean signal, s ̂ is the estimated value of the clean signal using some procedure. The term $\sum_{i=1}^n\frac{\partial\hat{s}}{\partial{s_i}}$ indicates the degrees of freedom of the estimator. In the case where the estimator can be written $y=Αx$, this term is equal to trace(A). 

\paragraph{Proposition 3}
For a signal of length L and denoised at N levels of decomposition,

\begin{multline}
\sum_{i=1}^n\frac{\partial\hat{s}}{\partial{s_i}}=trace(A)=\sum_{x=1}^L\delta_1(x)\max(\psi'_1*\psi_1) \\ 
+\sum_{j=1}^N\left(\sum_{x=1}^L\delta_j(x)\right) \max \left(\prod_{i=1}^{j-1}(\phi'_i*\phi_i)*(\psi'_j*\psi_j)\right) 
+L\times \max\left(\prod_{j=1}^N(\phi'_j*\phi_j)\right)
\end{multline}

Proof:
In the wavelet denoising scheme given by (13), we treat wavelet denoising as a convolution operation
$$\hat{s}(x)=A*s(x)+B $$
Convolution can be written as a matrix operation s ̂=Αs+Β where Α is the matrix whose diagonal contains kernels $\alpha_x$, and zeros elsewhere:

$$
\hat{s}=As =
\left[ \begin{array}{ccccc}
\alpha_1&0&0&\hdots&0 \\
0 & 	\alpha_2 &0 &\hdots & 0 \\
0 & 0 &\alpha_3 &\hdots&0\\
\vdots & & & &\vdots\\
0&0&\vdots&\vdots&\alpha_L\\
\end{array} \right]
\left[
\begin{array}{c}
s_1\\s_2\\s_3\\\vdots\\s_L
\end{array}
\right]
$$

$$
\frac{\partial \hat{s}}{\partial s}=A
$$

$$
\sum_{i=1}^n\frac{\partial\hat{s}}{\partial s_x}=\sum_{x=1}^{n}\max(\alpha_x)=trace(A)
$$

where
$$
\alpha_x=\delta_{1x}(\psi'_1*\psi)+\sum_{j=1}^N\delta_{jx}\prod_{i=1}^{j-1}(\phi'_i*\phi_i*)*(\psi'_j*\psi_j)+\prod_{j=1}^N(\phi'_j*\phi_j)
$$

The trace of the matrix $A$, the sum of the diagonal, is the sum of the amplitude of the centre of the kernel $\alpha_x$.

Therefore, for a signal of length L and denoised at N levels of decomposition,

$$
\sum_{i=1}^n\frac{\partial\hat{s}}{\partial{s_i}}=\sum_{x=1}^L\delta_1(x)\max(\psi'_1*\psi_1) 
+\sum_{j=1}^N\left(\sum_{x=1}^L\delta_j(x)\right) \max \left(\prod_{i=1}^{j-1}(\phi'_i*\phi_i)*(\psi'_j*\psi_j)\right) 
$$
$$
+L\times \max\left(\prod_{j=1}^N(\phi'_j*\phi_j)\right)
$$

In this work, we will use SURE to determine N, the number of levels of wavelet decomposition to be used in denoising using Stationary Wavelet Transform and the Universal Threshold thresholding rule. This parameter will depend on the length of the signal i.e. the number of data points and also the level of noise. Generally longer and noisier signals require more levels of wavelet decomposition.

As the optimum number is reached, the risk is expected to reach minimum before rising again with successively larger number of levels, signifying oversmoothing.

The value of the risk will depend on the value of the noise variance $\sigma^2$ used in the calculation. We will examine two cases, first where we use the known value of $\sigma^2$, and then a value estimated from the first level wavelet detail coefficients, via the median absolute deviation. The estimated noise variance is determined using the first level detail coefficients, calculating the Median Absolute Deviation.

\section{Methodology}
The wavelet denoising procedure is implemented using Scilab 5.4.1, an open source mathematics software. 
We use a series function 
\begin{equation}
\sum_{n=1}^M-\frac{10x}{n}\cos(a n x)
\label{eq: test signal}
\end{equation}

which is intended to mimic the absorption spectrum of water vapour.  The constant a determines the number of troughs. Figure 2, shows the signal with a = 3, 6, 9 and 12, from top to bottom, offset by 6. The signal contains large troughs and smaller features between the troughs. Additive White Gaussian Noise (AWGN) with known noise variance is added to the signal. We limit the number of iterations M to 18.

\paragraph{}

\begin{table}[H]
\caption{Signals generated by different values of $a$}
\begin{center}
\begin{tabular}{c | c c c c }
\textbf{$a$} & 3 & 6 & 9 & 12 \\
{Signal} & 1 & 2 & 3 & 4\\
\end{tabular}
\end{center}
\end{table}

\begin{figure}[H]
\begin{center}

\includegraphics[width=360pt]{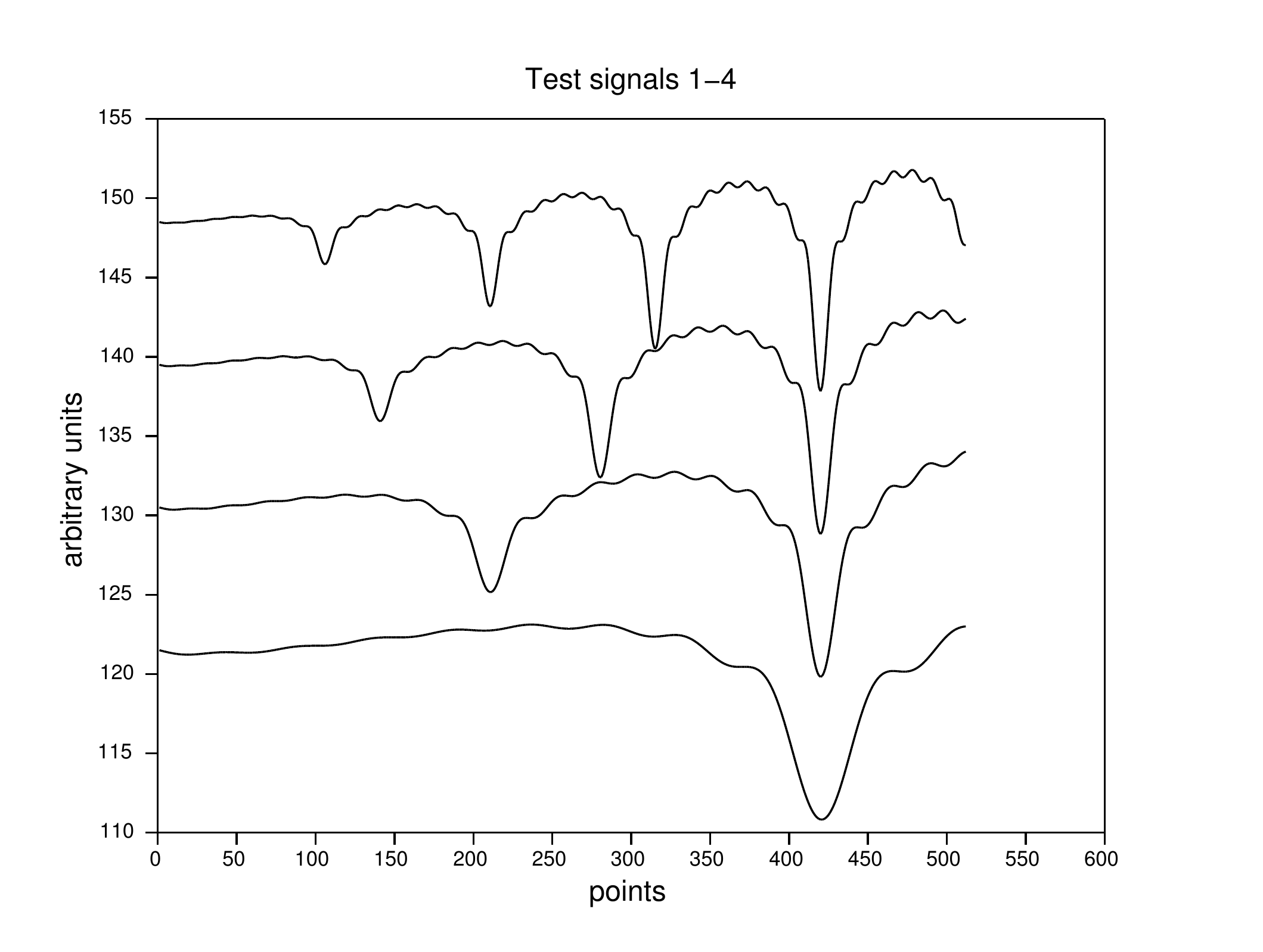}
\caption{From bottom to top: test signals 1-4 generated using \ref{eq: test signal} with a= 3, 6, 9  and 12.}
\end{center}
\end{figure}

The signal is denoised using the SWT with the VisuShrink soft thresholding rule for several levels of decomposition. 

We calculate the following values for each level of decomposition:
\begin{itemize}
	\item Sum square error between the known clean signal and the denoised signal
	\item SURE risk using noise variance known a priori 
	\item SURE risk using an estimate of the noise variance
\end{itemize}

The average value of thirty trials is taken for each method, for each level of decomposition.

Since the value of the SURE risk is dependent on the value of the noise variance used in the equation, we calculate the risk using both the a priori variance and also using an estimate. The estimate is calculated using the first level wavelet detail coefficients, with the assumption that is consists of noise. The sum square error and both cases of the SURE risk are plotted against level of decomposition to determine the level which obtain the least value, which indicate the optimum level of decomposition. The sum square error result is compared to the SURE risk.

The SURE risk value is not intended to substitute the sum square error and is not expected to have equal value to it. As a surrogate of MSE, SURE is expected to attain minimum value at the same level of decomposition as MSE. For the wavelet denoising procedure, we will use the Daubechies 4 (db4) wavelet and the Universal Threshold rule, with the Stationary Wavelet Transform. The test signal is inoculated with additive Gaussian white noise with noise variance $\sigma^2=0.25$.

\section{Results}

The results in Table 1 indicate that the SURE risk reaches minimum at the same level of decomposition as the sum square error, for both the case where the noise variance is known beforehand, as well as the noise variance estimated using the first level wavelet detail coefficients, despite the estimated noise variance differing from the actual variance by as much as 10\%. The method can be used to determine the optimum number of levels of decomposition to be used in denoising.

\begin{center}
\begin{tabular}{c | c c c c }
Signal & 1 & 2 & 3 & 4\\
\textbf{Method} &  \multicolumn{4}{c} {\textbf{Optimum Level}} \\ 

\hline

{Sum Square Error} & 4 & 3 & 3 & 2\\

{SURE Risk (a priori $\sigma^2$=0.25)} & 4 & 3 & 3 & 2\\

{SURE Risk (estimated $\sigma^2$)} & 4 & 3 & 3 & 2\\

\end{tabular}

\end{center}

The results can be seen in the following Figures, Figure 3-14.

\subsection{Signal 1}

\begin{figure}[H]
\includegraphics[width=300pt]{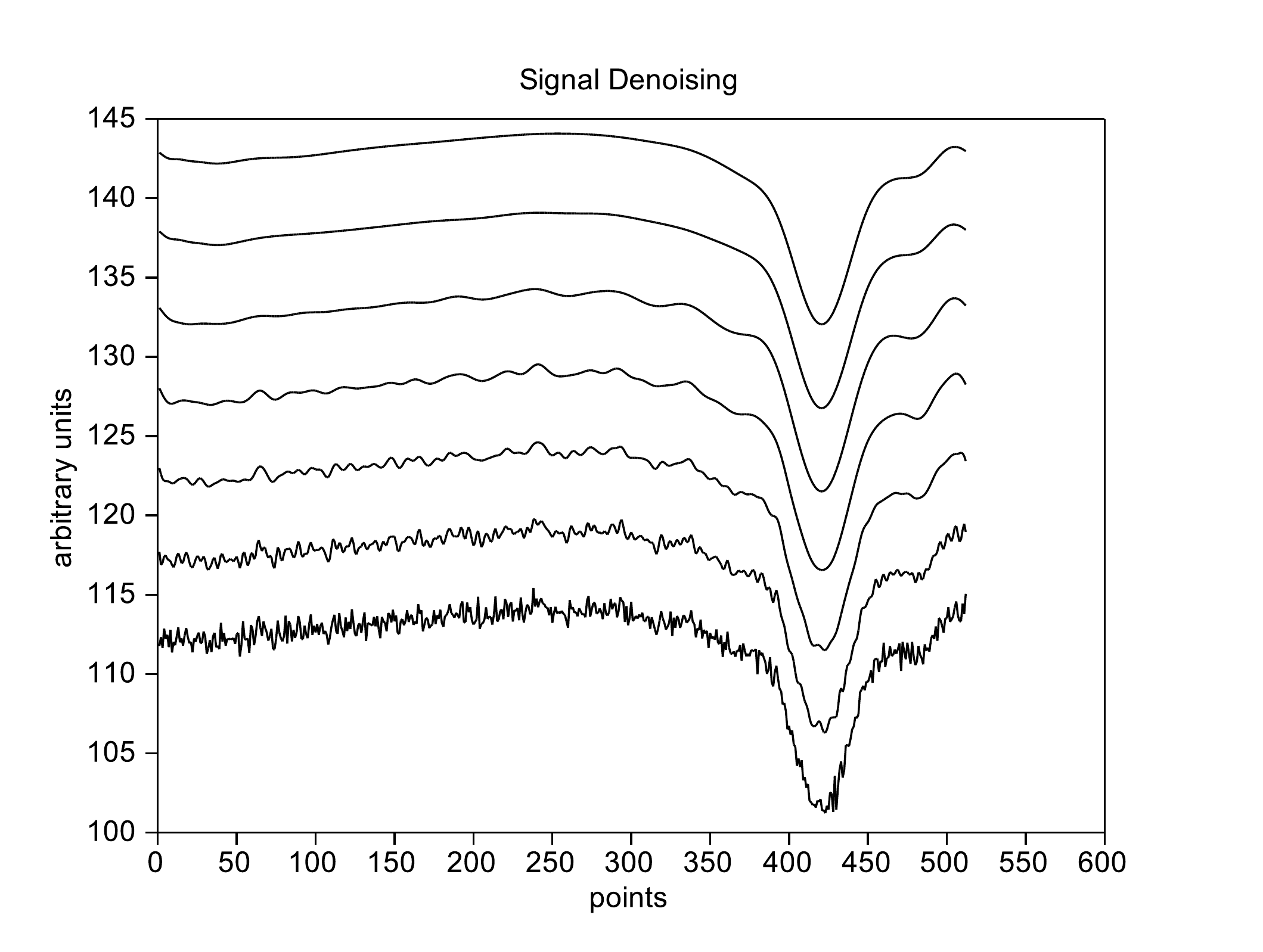}
\caption{Bottom to top: signal at 0-6 levels of decomposition}
\label{fig: signal 1}
\end{figure}

\begin{figure}[H]
\includegraphics[width=300pt]{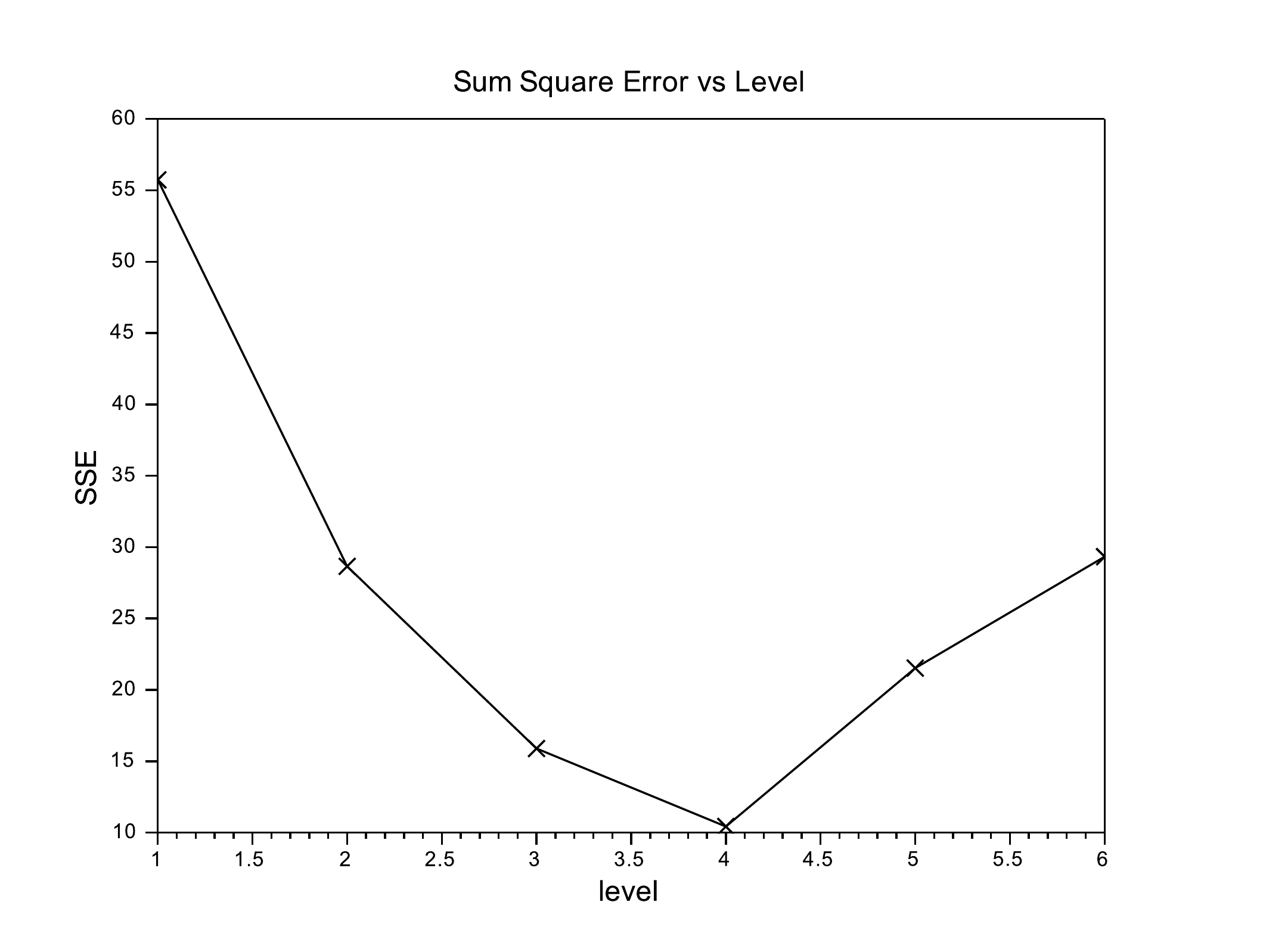}
\caption{Sum squared error}
\label{fig: signal 1 sse}
\end{figure}

\begin{figure}[H]
\includegraphics[width=300pt]{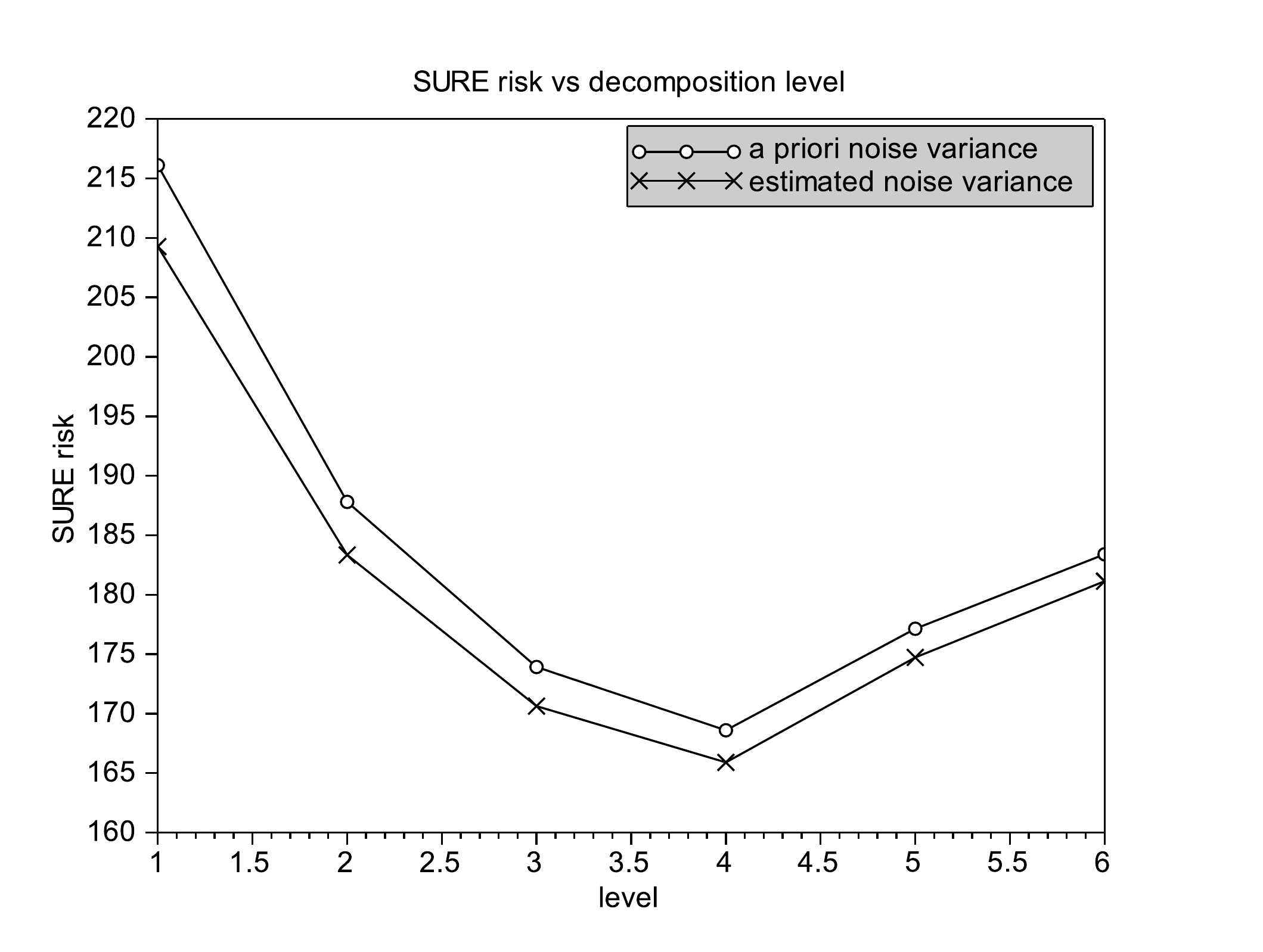}

\caption{Two estimates of the SURE risk}
\label{fig: signal 1 SURE}
\end{figure}

Figure \ref{fig: signal 1} shows the results for the first signal, for up to 6 levels of decomposition. Figure \ref{fig: signal 1 sse} shows the sum squared error, while Figure \ref{fig: signal 1 SURE} shows the values of SURE risk calculated using the known noise variance and the estimated noise variance respectively.


Figure \ref{fig: signal 1} shows the first signal being progressively denoised with successively higher levels of wavelet decomposition. The noisy signal is at the bottom, and the signals offset above it are the results of denoising at 1 to 6 levels of wavelet decomposition. As indicated in Table 2, the optimum level decomposition is 4 (the fifth signal from the bottom of Figure \ref{fig: signal 1}). At 5 levels of decomposition (sixth signal from bottom), we notice the loss of detail in the signal at $300< x <400$ as oversmoothing occurs.

\subsection{Signal 2}
\begin{figure}[H]
\includegraphics[width=300pt]{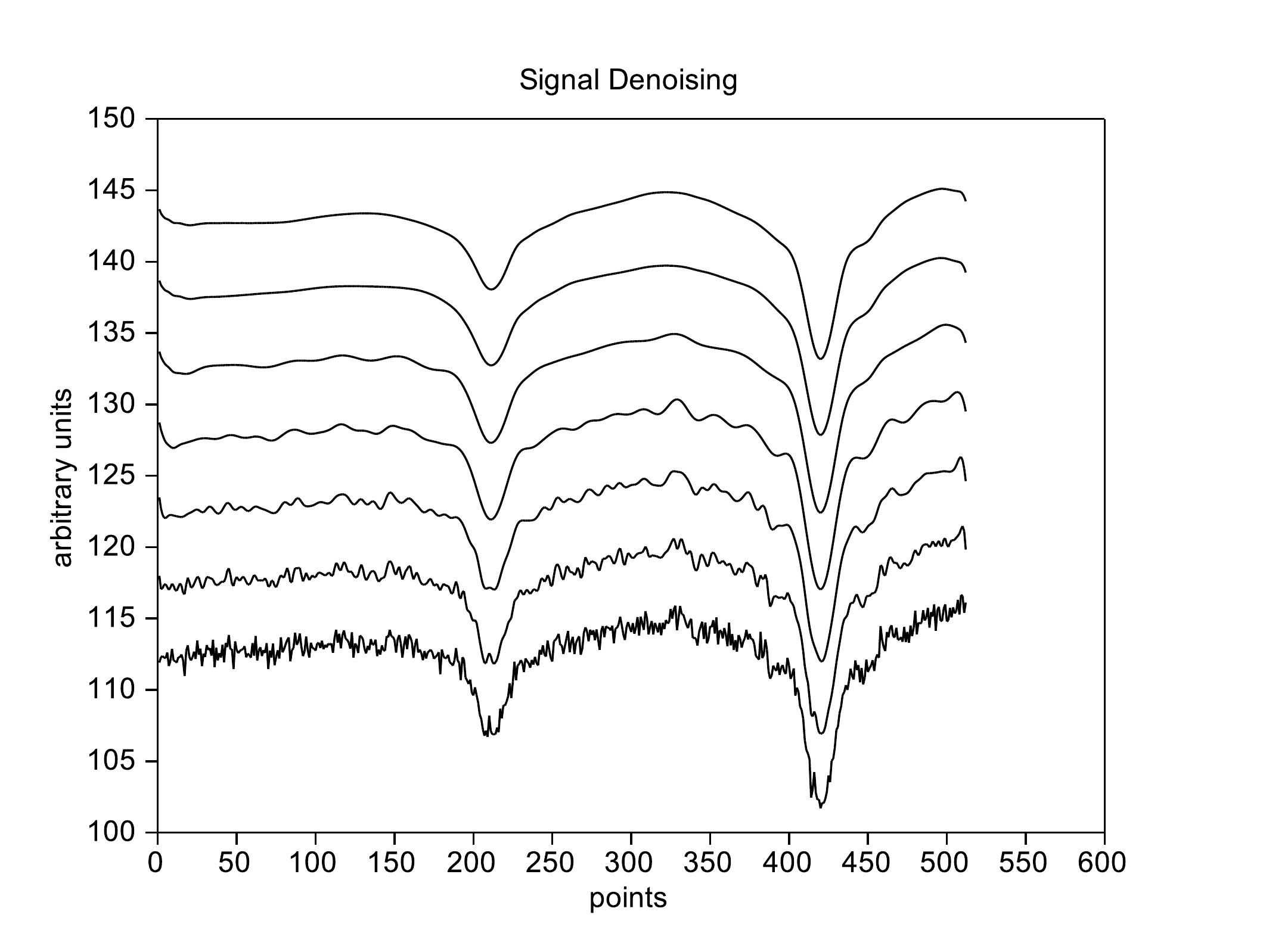}
\caption{Bottom to top: signal at 0-6 levels of decomposition}
\label{fig: signal 2}
\end{figure}

\begin{figure}[H]
\includegraphics[width=300pt]{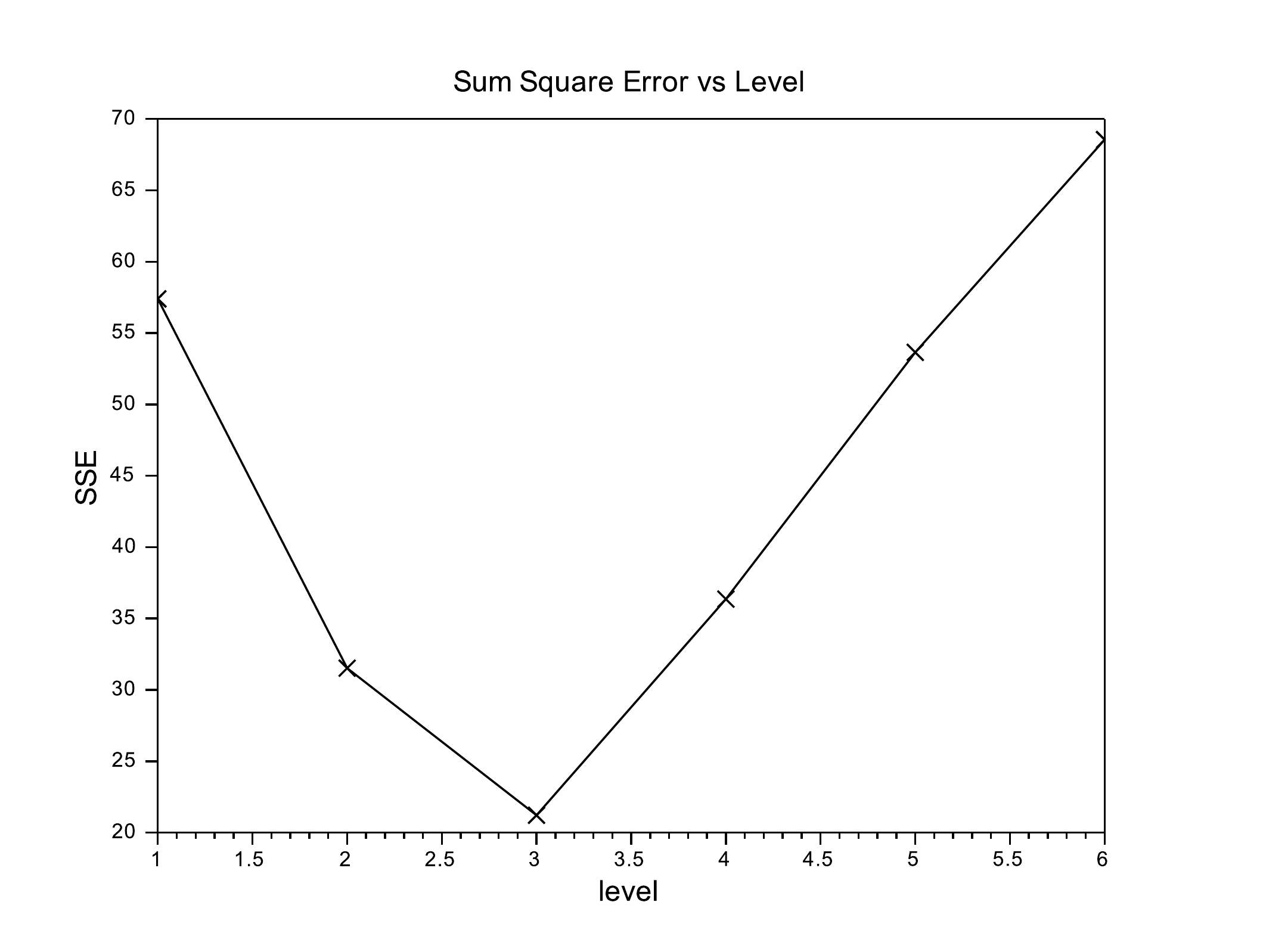}
\caption{Sum squared error}
\label{fig: signal 2 sse}
\end{figure}

\begin{figure}[H]
\includegraphics[width=300pt]{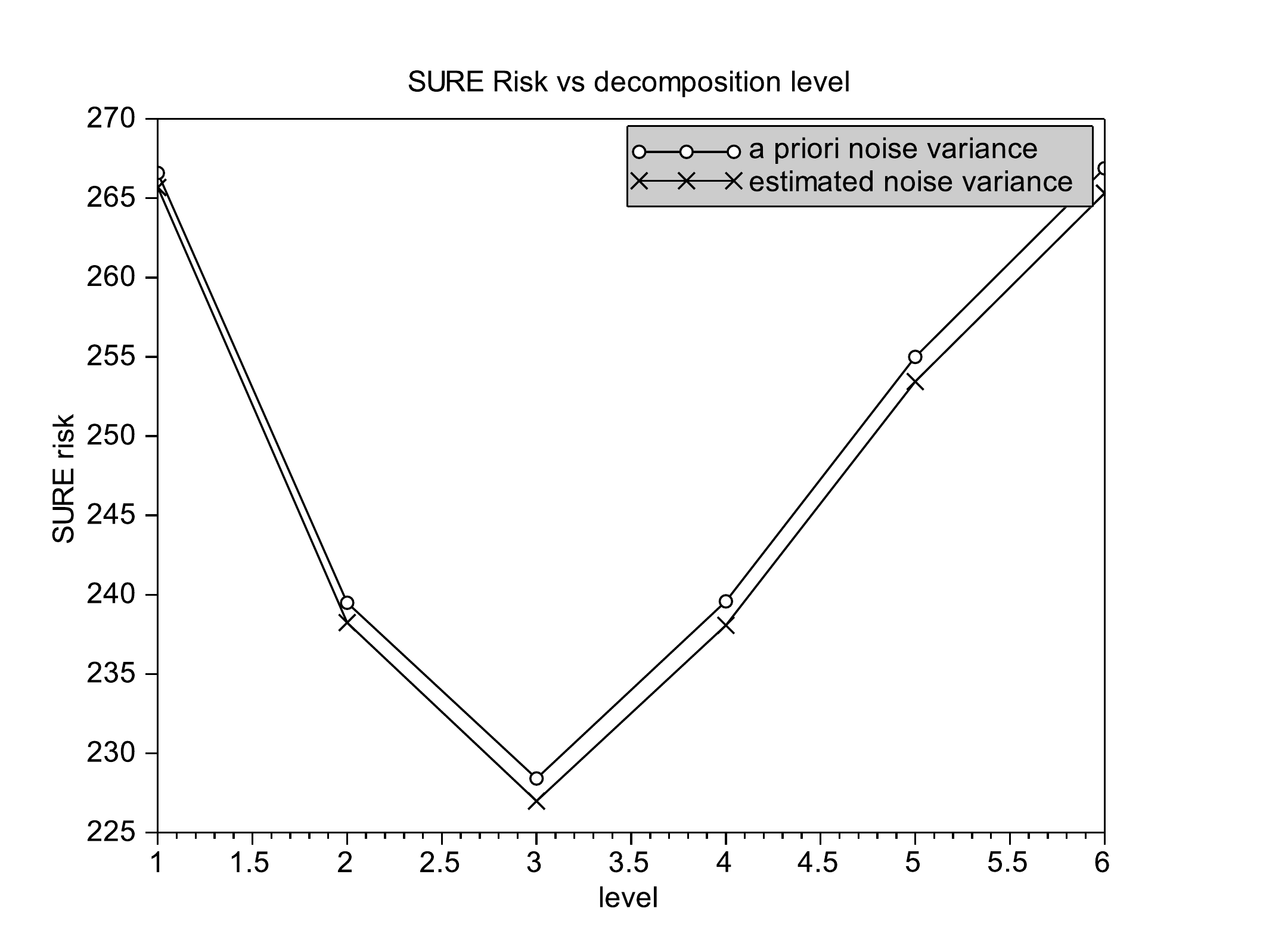}

\caption{Two estimates of the SURE risk}
\label{fig: signal 2 SURE}
\end{figure}

Figure \ref{fig: signal 2} shows the results for the second signal, for up to 6 levels of decomposition. Figure \ref{fig: signal 2 sse} shows the sum squared error, while Figure \ref{fig: signal 2 SURE} shows the values of SURE risk calculated using the known noise variance and the estimated noise variance respectively.

Figure \ref{fig: signal 2} shows the second signal being subjected to denoising. As before, the bottom signal is the noisy test signal. The signals above it show the result of denoising at successively higher levels of decomposition. At level 4 (fifth from the bottom) and above, the signal shows loss of detail. The smaller undulations between 250 and 400 outside the main peak are not preserved, as they are at the third level. At level 1 and 2, the signal still shows the presence of noise.

\subsection{Signal 3}
\begin{figure}[H]
\includegraphics[width=300pt]{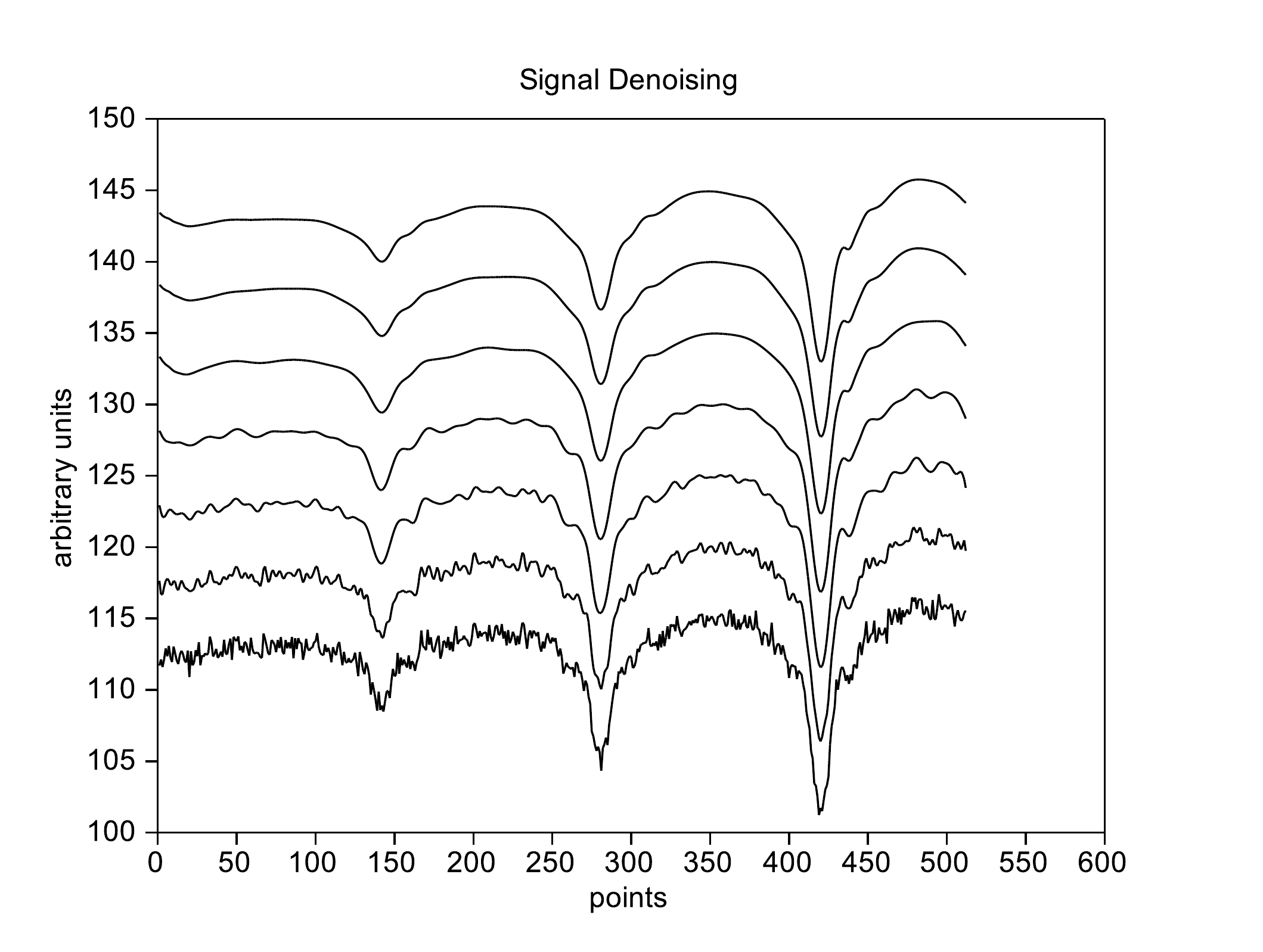}
\caption{Bottom to top: signal at 0-6 levels of decomposition}
\label{fig: signal 3}
\end{figure}

\begin{figure}[H]
\includegraphics[width=300pt]{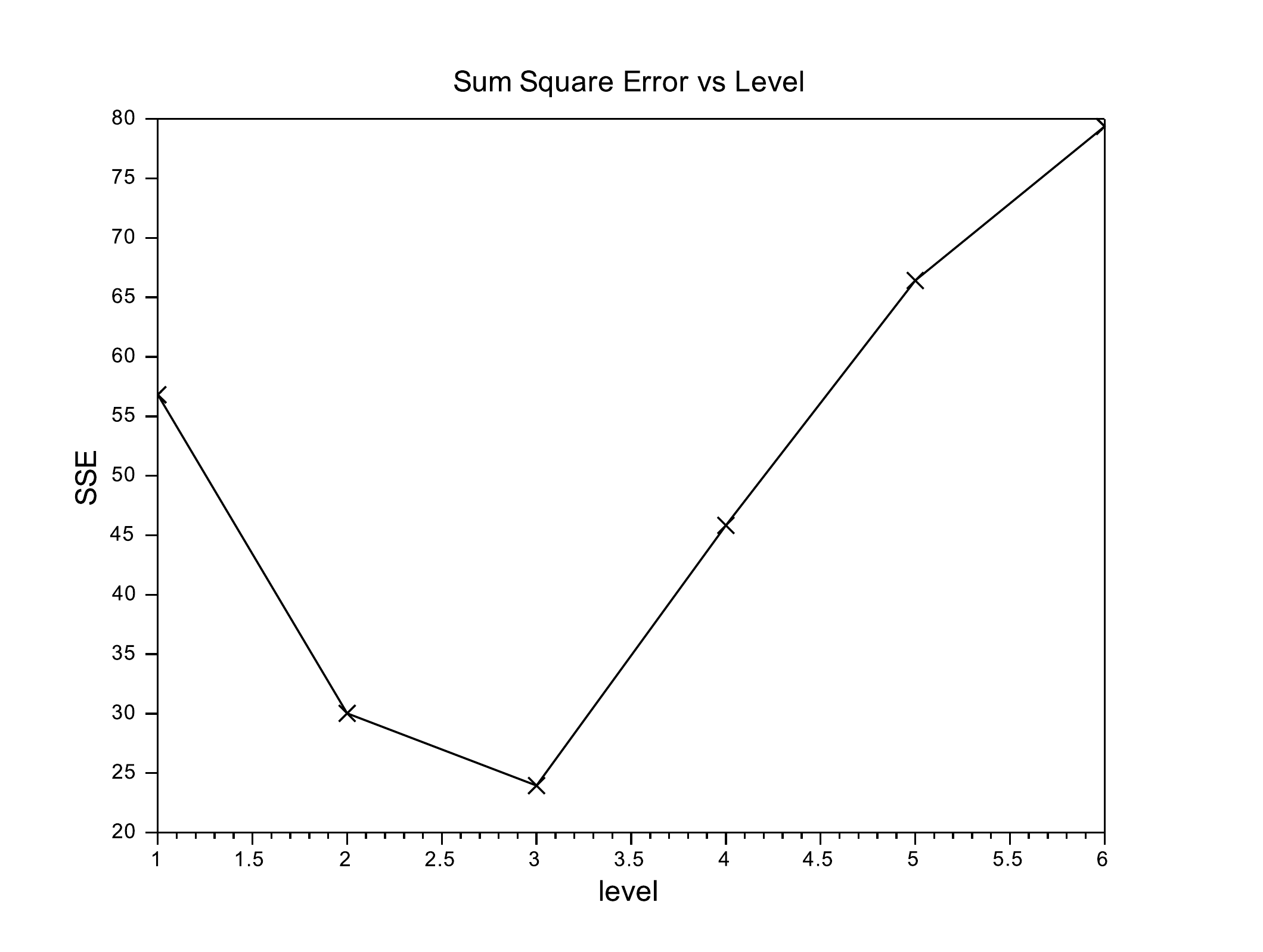}
\caption{Sum squared error}
\label{fig: signal 3 sse}
\end{figure}

\begin{figure}[H]
\includegraphics[width=300pt]{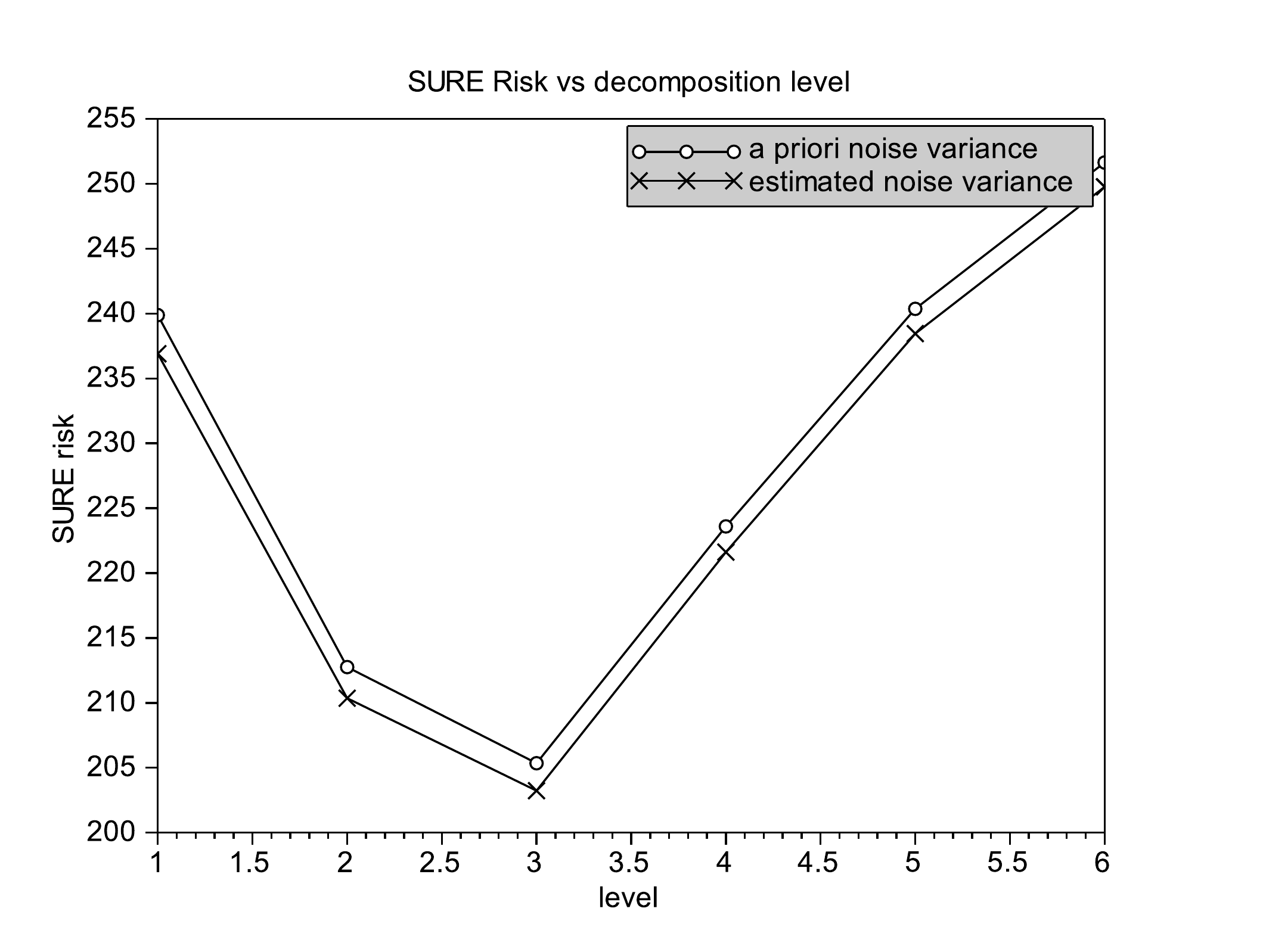}

\caption{Two estimates of the SURE risk}
\label{fig: signal 3 SURE}
\end{figure}

Figure \ref{fig: signal 3} shows the results for the third signal, for up to 6 levels of decomposition. Figure \ref{fig: signal 3 sse} shows the sum squared error, while Figure \ref{fig: signal 3 SURE} shows the values of SURE risk calculated using the known noise variance and the estimated noise variance respectively.

Figure \ref{fig: signal 3} shows the result of denoising at progressively higher levels of decomposition. As before, the bottom-most signal is the noisy signal and successively higher levels of decomposition are offset above it. The optimal level is three (fourth signal from bottom). The smaller features of the signal begin to disappear from the fourth level onwards (fifth signal from the bottom and up). 

\subsection{Signal 4}
\begin{figure}[H]
\includegraphics[width=300pt]{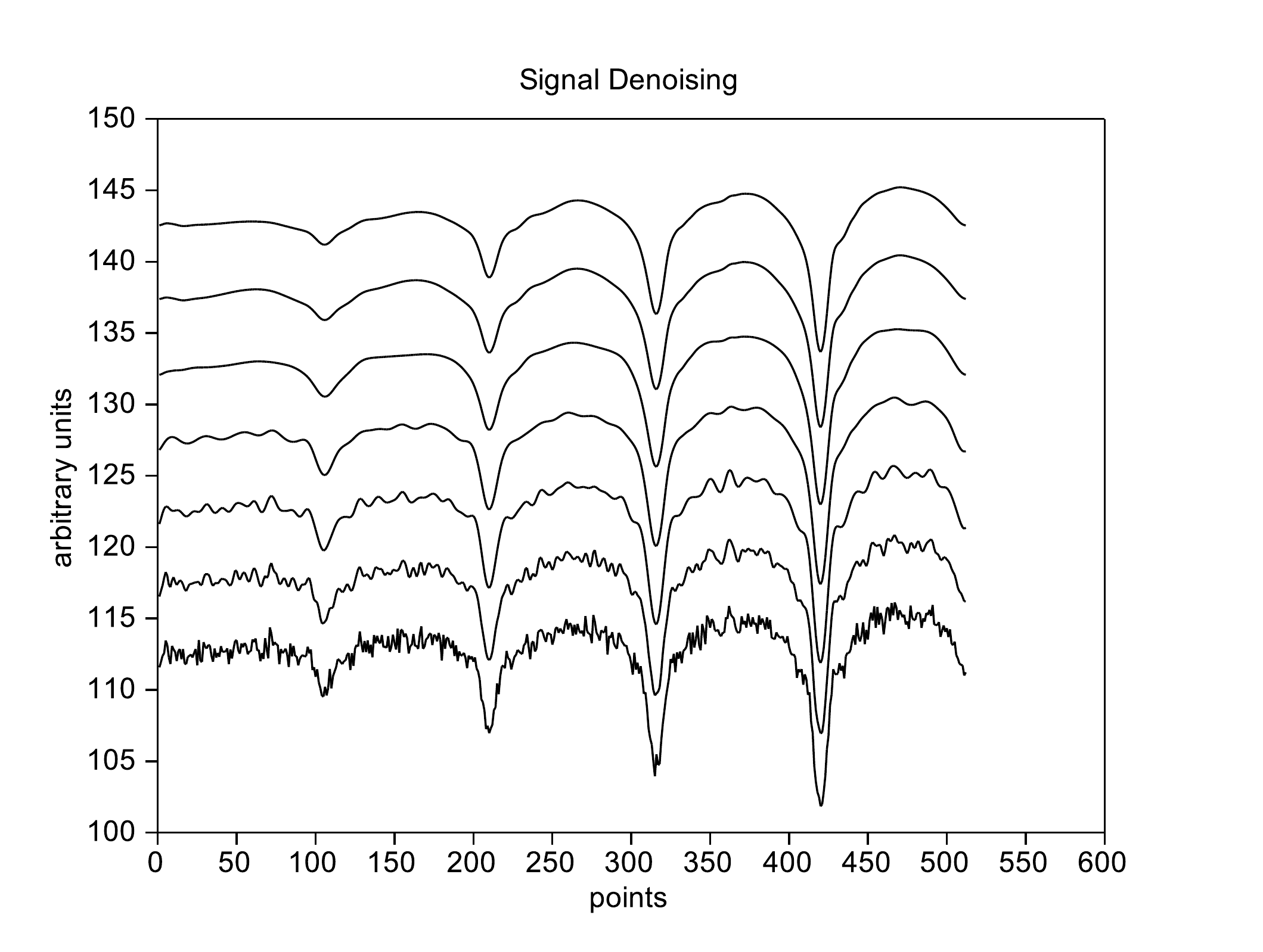}
\caption{Bottom to top: signal at 0-6 levels of decomposition}
\label{fig: signal 4}
\end{figure}

\begin{figure}[H]
\includegraphics[width=300pt]{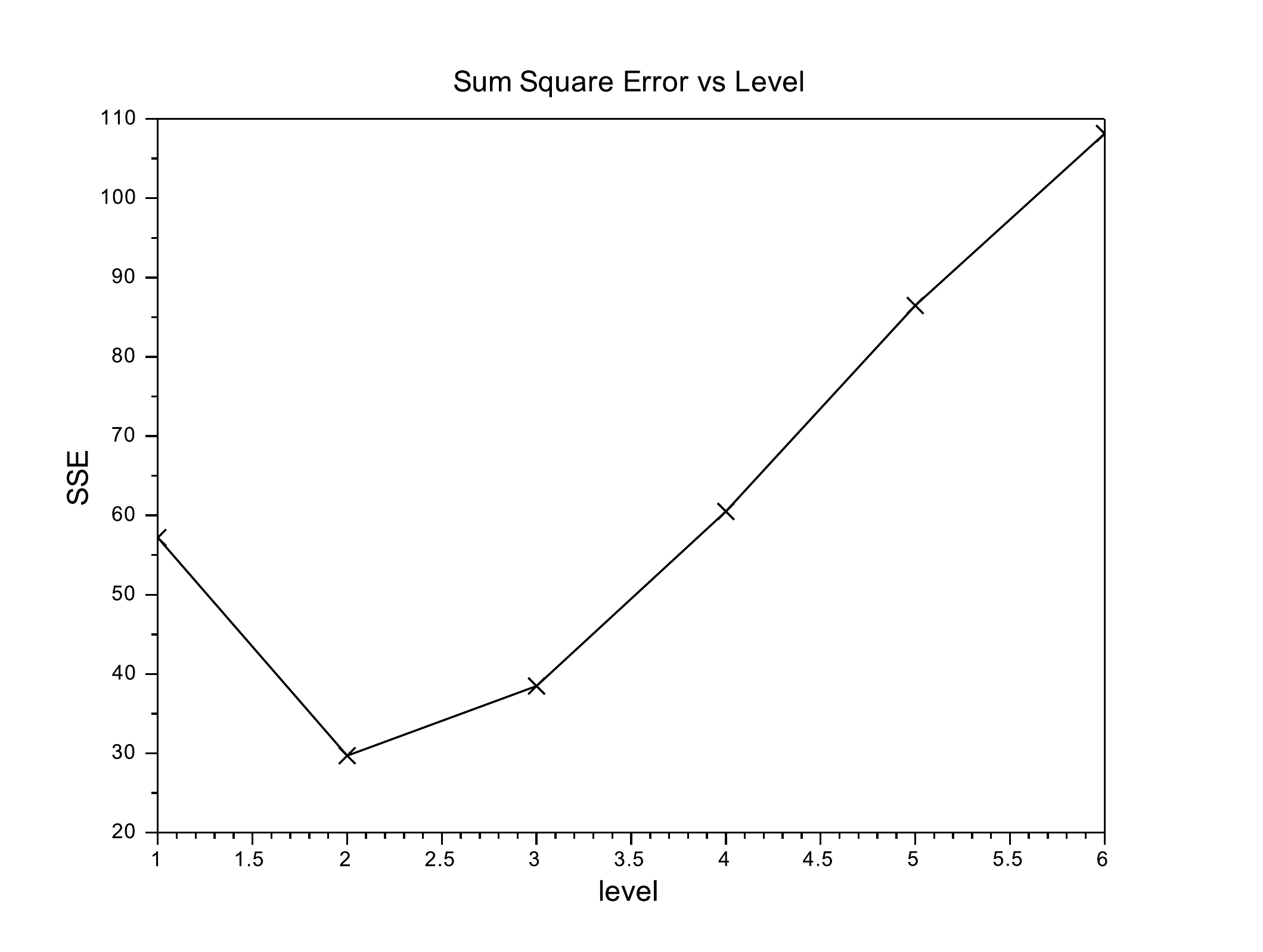}
\caption{Sum squared error}
\label{fig: signal 4 sse}
\end{figure}

\begin{figure}[H]
\includegraphics[width=300pt]{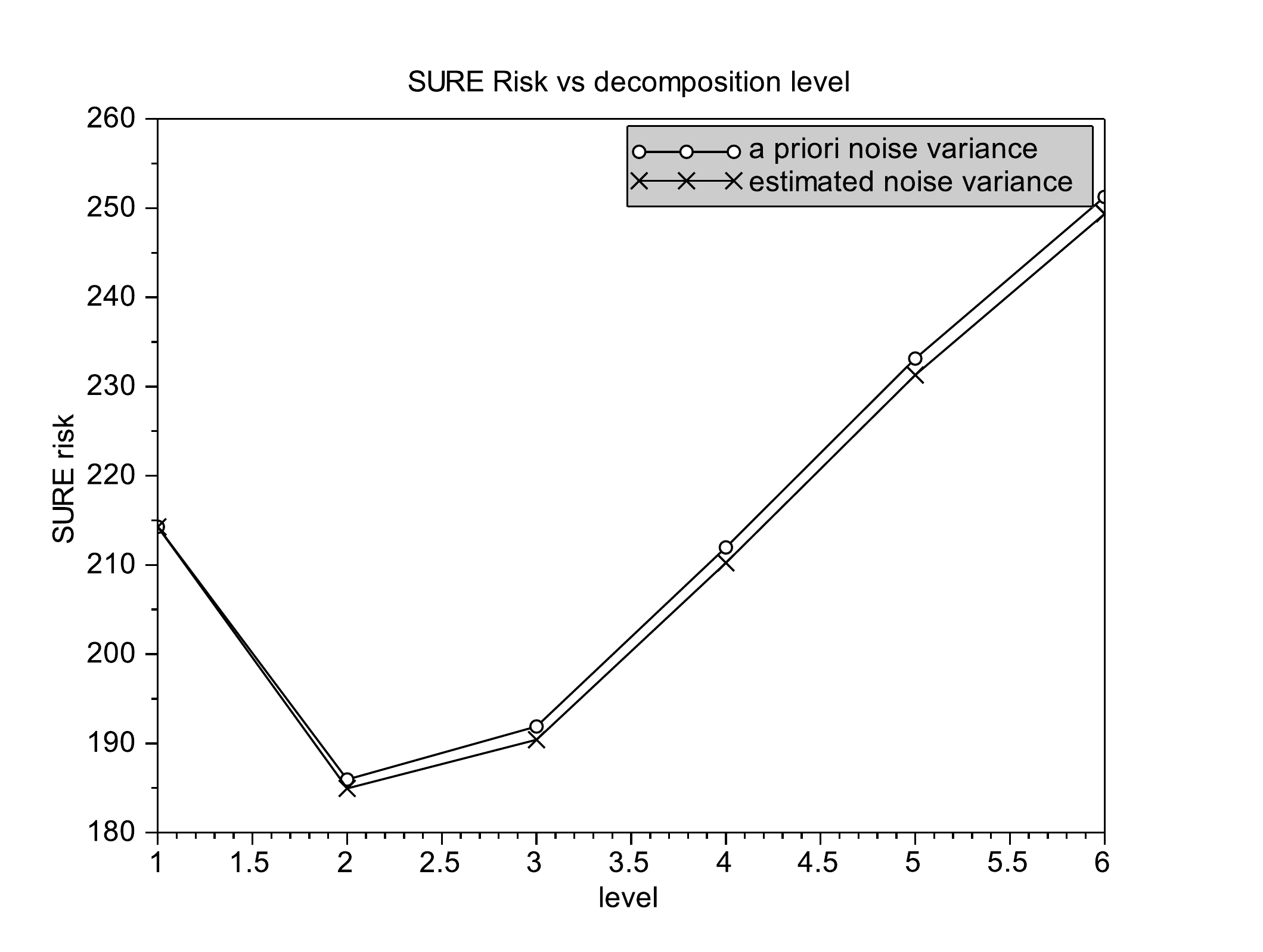}

\caption{Two estimates of the SURE risk}
\label{fig: signal 4 SURE}
\end{figure}

Figure \ref{fig: signal 4} shows the results for the fourth signal, for up to 6 levels of decomposition.  The bottom signal is the signal prior to denoising. At two levels of decomposition (thrid trace from the bottom-most), the signal is denoised while retaining the smaller features outside the large features of the signal. Figure \ref{fig: signal 4 sse} shows the sum squared error, while Figure \ref{fig: signal 4 SURE} shows the values of SURE risk calculated using the known noise variance and the estimated noise variance respectively.

\section{Conclusion}

A method of determining the optimum level of wavelet decomposition for wavelet denoising has been presented in this paper. The method requires the variance to be known in order to calculate the SURE risk associated with that level of decomposition. However, it has been shown that using an estimated value of the noise variance will yield similar results. The method has been found to indicate the same level of decomposition as using the sum-square error, and may be used for denoising a signal with unknown noise characteristics.

Further work that can be done include better estimation of the noise variance, upon which the calculation of the SURE risk estimate depends. The estimate of noise variance, which depends on the first level wavelet coefficients, assumes that the coefficients represent uniform noise. But under some circumstances, such as signals with discontinuities, the coefficients may contain outliers which can render the estimate unusable. The field of robust noise variance estimation will require study.

Other works that can follow include investigation of other wavelet thresholding rules, the use of other risk estimators rather than Stein's Lemma \cite{GubbiSeelamantula2014}, and the investigation of wavelet denoising in signals with Poisson noise statistics and mixed Gaussian-Poisson noise.

The algorithm can be optimized for speed by precalculating the heights of the kernels, since this is the only information that is actually used.

\section{References}

\bibliography{mybibfile}

\end{document}